\documentclass{aastex631}
\usepackage{amsmath}
\usepackage{amssymb}
\usepackage{hyperref}
\newcommand*\diff{\mathop{}\!\mathrm{d}}

\begin{document}

\title{Detection of Radio Emission from Super-flaring Solar-Type Stars in the VLA Sky Survey}
\shorttitle{Superflare Stars in VLASS}
\author{Ivey Davis}
\affiliation{Cahill Center for Astronomy and Astrophysics, MC 249-17 California Institute of Technology, Pasadena CA 91125, USA}

\author{Gregg Hallinan}
\affiliation{Cahill Center for Astronomy and Astrophysics, MC 249-17 California Institute of Technology, Pasadena CA 91125, USA}

\author{Carlos Ayala}
\affiliation{Cahill Center for Astronomy and Astrophysics, MC 249-17 California Institute of Technology, Pasadena CA 91125, USA}

\author{Dillon Dong} 
\affiliation{National Radio Astronomy Observatory, P.O. Box O, Socorro, NM 87801, USA}

\author{Steven Myers} 
\affiliation{National Radio Astronomy Observatory, P.O. Box O, Socorro, NM 87801, USA}

\begin{abstract}
Solar-type stars have been observed to flare at optical wavelengths to energies much higher than observed for the Sun. To date, no counterparts have been observed at longer wavelengths. We have searched the the VLA Sky Survey (VLASS) for radio emission associated with a sample of 150 single, solar-type stars previously been observed to exhibit superflares in the Transiting Exoplanet Survey Satellite (\emph{TESS}). Counterparts to six of these stars were present in VLASS as transient or highly variable radio sources. One of the stars is detected in all three epochs, exhibiting an unprecedented level of apparently persistent radio emission. The engine for this radio emission is unclear, but may be related to accretion, a binary companion, or the presence of large-scale magnetic field. Two stars show radio emission with $>50\%$ circular polarization fraction, indicating a coherent emission process likely being present. We find that the six VLASS-detected stars tend to have higher flare rates and higher flare energies of our \emph{TESS} sample. This, in addition to the VLASS-detected stars adhering to the G\"udel-Benz relation, suggest that the radio emission may be directly associated with superflares. These results confirm that the superflare phenomenon on solar-type stars extends to radio wavelengths, in this instance tracing particle acceleration. These data provide the first window on the luminosity function of radio superflares for solar-type stars and highlights the need for coordinated, multi-wavelength monitoring of such stars to fully illustrate the stellar flare-particle relation.
\end{abstract}

\section{Introduction}

Flares are the result of magnetic reconnection releasing energy stored in twisted field lines \citep{PriestForbes2002,ShibataMagara2011Review}. This energy goes into heating coronal plasma in the immediate vicinity of the reconnection site as well as into accelerating particles to produce non-thermal emission and to drive heating deeper in the atmosphere of a star. The various emission processes associated with flares, in addition to the various locales of the emissions' origins, makes it an intrinsically multi-wavelength phenomenon (see \citet{Kowalski2024} for a recent review). While the most energetic flares we see from the Sun are on the order of $\sim10^{32}$\,erg \citep{Woods2006}, other solar-type stars (\textit{i.e.} stars with temperatures between 5,600-6000\,K)  have been found to experience superflares-- flares in excess of $10^{33}$\,erg.

Superflares had been considered exceptionally rare \citep{Schaefer2000}, but the identification of nearly 400 superflares among $\sim150$ solar-type stars, including slowly-rotating stars (stars with periods greater than 10 days), in \emph{Kepler} data by \cite{Maehara2012} allowed for the beginning of a rigorous, statistical analysis of the occurrence rate of superflares on solar-type stars and solar analogs. Using a larger set of \emph{Kepler} data, \cite{Notsu2019} and \citet{Okamoto2021} estimate a slowly-rotating, solar-type star may have a $10^{34}$\,erg superflare every few thousand years and that this energy may represent the upper limit of the flare energy that a slowly-rotating star is able to produce. These results are roughly supported by earlier, first-principles calculations by \cite{Shibata2013}. Alternatively, solar-type stars with periods less than a few days can produce flares with orders of magnitude higher energy and much more frequently \citep{Davenport2016, Tu2020, Doyle2020}.

These results have important implications when considering possible associated particle escape in the form of coronal mass ejections (CMEs) and solar/stellar energetic particle (SEP) events, especially in the context of habitability and abiogenesis \citep{Airapetian2016, Airapetian2020}. Stellar CMEs are notoriously difficult to unambiguously detect \citep{Namekata2022}. As such, there are no empirical relations for correlating stellar flares with stellar CME occurrence. On the sun, more energetic flares are more likely to have an associated CME and \citep{Yashiro2009} and more energetic flares are related to more energetic CMEs \citep{Emslie2012, Drake2013}. It might then be expected that younger, more active stars with more energetic flares would also have more energetic CMEs \citep{Aarnio2012} and that this may have important implications for both the rotational evolution of the Sun \citep{Skumanich1972, Xu2024} and the particle environment that a young earth might have been subjected to. Alternatively, relationships between solar flares and CMEs may break down for active stars with stronger, large-scale magnetic fields which may suppress mass escape \citep{AlvaradoGomez2022}. 

Radio emission at frequencies below a few GHz can provide a window into particle acceleration in stellar coronae. This is supported by the G\"udel-Benz relationship which shows that the quiescent radio luminosity of binaries and G, K, and M stars have a correlation with their quiescent soft X-ray luminosity \citep{GudelBenz1993}. The G\"udel-Benz relationship also holds for flare-related emission both from stars and from the Sun \citep{BenzGudel1994}. The correlation between the non-thermal radio emission (driven by electrons trapped in closed magnetic field loops) and the thermal X-ray emission (produced by the hot coronal plasma) has been used as evidence that the corona is heated via flare-related processes. This relationship can break down when the emission process is coherent-- either plasma emission or electron cyclotron maser emission (ECME). This occurs on the Sun for the most extreme radio burst events, the properties of which may be indicative of a fundamental shift in the conditions and acceleration processes responsible for producing radio bursts \citep{Gary2019,Cliver2022}. Although radio emission associated with flare activity has been observed for especially magnetic active stars such as UV Ceti variables \citep{Spangler1976, Osten2005, Zic2020}, there has been no observed radio counterpart to the stellar superflare phenomenon on solar-type stars. Because of this, it remains unclear whether the G\"udel-Benz relationship holds for these events, if there's a divorce from the relationship that could be indicative of a transition in emission process for highly energetic events, and what this might mean for the type of particle acceleration occurring and its impact on the surrounding stellar environment.

Flare occurrence is stochastic, and guaranteeing a detection of a superflare from even an incredibly active star can require days of observing time dedicated to that star. Without dedicated, multi-wavelength monitoring of stars, the detection of the radio analogs to the superflare phenomenon relies on wide-field survey data. This has been partially accommodated by radio surveys like the Low-Frequency Array (LOFAR) Two-metre Sky Survey (LoTSS) \citep{Shimwell2017, Yiu2023},  surveys from the Australian Square Kilometer Array Pathfinder (ASKAP) \citep{Pritchard2021, Pritchard2024}, and the Very Large Array Sky Survey (VLASS) \citep{Lacy2020}.  Ayala et al. (in preparation) have recently identified $\sim80$ radio transients associated with stellar counterparts. Two of these stars are young, solar-type stars found by \citet{Tu2020} and \citet{Doyle2020} to have superflares as observed by the Transiting Exoplanet Survey Satellite (\emph{TESS}). In this paper, we expand the search for radio emission associated with super-flaring solar-type stars by extending the search to two epochs and optimized search criteria that rely on an existing catalog of known \emph{TESS} superflare stars.  We use the work of \citet{Tu2020} and \citet{Doyle2020}-- both of whom focused on identifying flares from solar-type stars in the first year of \emph{TESS} observations-- to build a list of 150 stars to search for radio emission in VLASS as described in Sections \ref{sec:data} and \ref{sec:methods}. Of these stars, we find radio emission from four additional stars, bringing the total number of \emph{TESS} super-flaring stars with a VLASS counterpart to six. A summary of these stars is provided in Table \ref{tab:flare_star_info}.

\begin{table*}[ht]
\centering
\begin{tabular}{lcccccccccc}
\hline \hline
Star &TIC &Association &$d$ &$R$ &$T_\text{eff}$ &$P$ &Ro &Max Energy &$N_\text{fl}$& Flare rate \\
& & &pc &R$_\odot$ &K &d & &$\log_{10}(\text{erg})$ & &yr$^{-1}$ \\\hline
HD 22213 &93122097 & THA &51.378 &1.01 &5456 &1.393 &0.0292 &34.64 & 9 &77 \\
HD 295290 &53417036 & Col &60.526 &0.98 &5554 &1.522 &0.0295 &35.73 & 26 & 217 \\
AT Col &144499196 & Field &75.802 &1.26 &5291 &2.507 &0.0479 &36.08 & 40 &159 \\
HD 245567 &52588257 & Field &106.533 &1.53 &5362 &1.498 &0.0301 &35.50 & 14 & 59 \\
HD 156097 &152346470 &Field &115.504 &1.44 &5844 &1.976 &0.043 &35.92 & 31 &246 \\
HD 321958 &79358659 &UCL &172.765 &1.26 &5626 &1.466 &0.0289 &36.41 & 16 &130 \\
Median Star & --- & --- & 121.759 & 1.05 & 5528 & 2.027 & 0.0387 & 35.41 & 6 & 56  \\
\hline
\end{tabular}
\caption{Information on the six \emph{TESS} super-flaring stars found in VLASS. The associations in the third column have certainties $>70\%$ as calculated by Banyan \citep{Gagne2018}. Here, $d$ is the distance to the star, $R$ is the stellar radius reported by \citet{GaiaCollaboration2018}, and $T_{\text{eff}}$ is the effective stellar temperature as reported by \citet{Gaia2022}. $P$ is the period we found from the box-least-squares periodogram of the \emph{TESS} data from the \texttt{lightkurve} package. Ro is the Rossby number calculated using equation 11 in \citet{Cosaro2021}. The Max Energy is the largest \emph{TESS} flare energy we calculated and $N_\text{fl}$ is the number of \emph{TESS} flares we identified for a star. The energy, flare number, and flare rate calculations are all described in Appendix \ref{app:flare_characterization}. The Median Star row refers to the median value for the sample of 150 \emph{TESS} super-flaring stars. \label{tab:flare_star_info}} 
\end{table*}

\section{Data and Selection}\label{sec:data}
Because we are interested in the radio counterpart to the superflare phenomenon in solar-type stars and investigating whether the radio emission is similarly prevalent, the sample of stars we search for emission in VLASS were those identified to have flares in the catalogs of \citet{Tu2020} and \citet{Doyle2020}. Both of these catalogs focus on solar-type stars during the first year of \emph{TESS} data. This is motivated by the stars that showed up in Ayala et al. (in preparation) that were also present in these catalogs (HD 245567 and HD 156097). A summary of the data and source selection criteria are provided in this section.

\subsection{VLASS}

With a survey speed of $\sim 23.83$ square degrees per hour to a depth of $\sim120\mu$Jy/beam, VLASS presents the 2-4\,GHz sky above $-40^\circ$ declination to a sensitivity of  for each of three epochs \citep{Lacy2020} spaced by $\approx32\,$ months. This survey is being conducted in the B and BnA configurations of the VLA, providing a  spatial resolution of $\sim2.5"$. At the time of this paper, VLASS has completed the first two epochs and first half of the third epoch. The combination of spatial resolution, large sky coverage, and multi-epoch nature of VLASS has made it an exemplary resource for looking for radio transients \citep{Dong2021, Dong2023, Somalwar2023}.

Ayala et al. (in preparation)'s search for stellar radio emission identified two solar-type stars previously found to produce superflares \citep{Doyle2020, Tu2020}. There are ~$10^{11}$ synthesized beams in each epoch of VLASS and a transient search in the full survey area requires a stringent detection threshold to avoid false positives due to thermal noise fluctuations. Because of this, the transient-identification requirements chosen were $7\sigma$ in E2 and $<3\sigma$ detection in E1 where $\sigma$ is the noise of the field. We are searching for VLASS sources associated with a pre-existing sample of only a few hundred flaring stars. Because of this, our detection threshold is much less strict; we use the requirement that candidate sources are coincident with the location of a flaring star and are $\geq4\sigma$ in either epoch (see Section \ref{subsec:vlass_method} for more details). 

\subsection{Flaring Solar-Type Stars in \emph{TESS}}
\emph{TESS}'s photometric precision for bright targets ($\sim70$\,parts-per-million (ppm) for 1\,hr integration in the $I_C$ band), huge field of view ($24^\circ\times 96^\circ$) \citep{Ricker2014}, and persistent observing has made it an exceptional tool to expand our understanding of flares and the kinds of stars that produce them (\textit{e.g.} \cite{Gunther2020,Tu2020,Doyle2020,Pietras2022,Feinstein2022}). \emph{TESS}'s red (600-1000\,nm) bandpass was chosen to optimize precision of transits around M-dwarfs. Because flares peak in the bluer part of the spectrum than the stellar surface, and because the photosphere of solar-type stars are larger than M-dwarfs, the contrast of flares on solar-type stars is much lower than for flares on M-dwarfs; generally, \emph{TESS} can only detect extremely high-energy flares on solar-type stars. Although \emph{Kepler}'s bluer bandpass is sensitive to a broader range of flare energies, it covers $<1\%$ of the sky covered by VLASS \citep{Koch2010, Lacy2020}; \emph{TESS} has a much larger sky overlap with VLASS, even from just \emph{TESS}'s first year of observing. \cite{Tu2020} and \cite{Doyle2020} have used the \emph{TESS} 2-minute cadence light curves to analyze flares from solar-type stars observed by \emph{TESS}'s first 13 sectors. From this data, \citet{Tu2020} identified 1216 superflares on 400 stars and \citet{Doyle2020} find 1980 flares from 209 stars.

Of the 400 stars in the \citet{Tu2020} sample, only 112 also showed up in the \citet{Doyle2020} collection. The discrepancy in the sources that show up in the catalogs is due to the two studies having slightly different selection criteria for solar-type stars as well as flare identification procedures and methodologies for classifying and reporting flare information. For instance, \citet{Tu2020} exclude all targets that had a bright star within $42"$ (two pixels on the \emph{TESS} detector) of its position in order to avoid contaminated light curves. They have also attempted to remove binaries from their selection as identified in the \textit{Hipparcos-2} catalog \citep{vanLeeuwen2007}. \citet{Doyle2020} implemented neither of these restrictions. Additionally, while \citet{Doyle2020} accounted for the \emph{TESS} band when calculating the flux from the stars, neither \citet{Doyle2020} nor \citet{Tu2020} accounted for the difference in color between the flare and the quiescent stellar surface and what this would mean for their relative flux contribution in the \emph{TESS} band. 

Although the actual stellar flare temperature can be much higher than the $\approx9000\,$K that is often assumed in flare energy analyses \citep{Berger2023}, excluding any temperature difference between the photosphere and flare leads to underestimating the bolometric energy of the flare. This likely especially affects \citet{Tu2020}'s catalog, which focused specifically on superflares and so required flares be $>10^{33}\,$erg to be included in their catalog. See Appendix \ref{app:colors} for a deeper discussion of this underestimate.

The first year of \emph{TESS}'s operations focused on the southern hemisphere and had a maximum declination of $\approx15^\circ$. Because VLASS has a minimum declination of $-40^\circ$, only $\approx55\%$ of the sky covered by VLASS had been covered by \emph{TESS} when \citet{Tu2020} and \citet{Doyle2020} produced their superflare star catalogs. 74 of the 209 flaring stars reported in \citet{Doyle2020} and 146 of the 400 flaring stars reported by \citet{Tu2020} were in a field covered by VLASS. Accounting for stars that showed up in both catalogs, there was a total of 180 stars between the two catalogs that were in fields covered by VLASS. 

\subsection{Addressing Binarity}\label{subsec:binarity}
For isolated, solar-type stars, the spin evolution of the star is driven by magnetic braking, wherein angular momentum is lost to stellar winds and mass ejections interacting with the stellar magnetic field \citep{Skumanich1972}. In binary systems, tidal interactions can also modify the rotational evolution, resulting in rapid rotation for compact binaries. Gyrochronology is moot as an age tracer in these systems. Examples include RS CVn \citep{Hall1976} and some Algol-type systems \citep{Richards1992}. Because we are interested in stars that may be representative of a young solar system, we want to exclude any systems whose histories or activity may be significantly affected by a stellar companion. 

Work by \citet{Fleming2019} suggests that low-mass stars ($M \lesssim M_\odot$) can tidally interact out to orbital periods of 100\,days. This is roughly consistent with synchronous timescales described in \citet{Zahn1977}, assuming a second degree Love number $k_2 \approx 0.5$ \citep{Fleming2019} and a ratio between the stellar moment of inertia to the moment of inertia of a ring of $\approx0.07$ \citep{Claret1989}. For binaries composed of solar-mass stars, this then requires that the stars be separated by $\gtrsim0.5\,$AU.

For all 180 stars from \citet{Tu2020} and \citet{Doyle2020}'s catalogs that were covered in a VLASS field, we checked the classification as listed on SIMBAD \citep{Wenger2000} to identify RS CVns. We also checked the Gaia DR3 non-single star (NSS) catalog \citep{Gaia2022}, the Washington Double Star (WDS) catalog \citep{Mason2001}, and the spectroscopic binary orbit catalog \citep{Pourbaix2004} in order to identify as many binaries as possible where the orbit may have played a significant role in the rotational properties of the star. Any star that was reported by \citet{Gaia2022} or \citet{Pourbaix2004} to have an orbital period $<100\,$days was flagged as a binary. Additionally, we required that the quality code provided for the binary in \citet{Pourbaix2004} be 3 or higher. We used the largest angular separation between physical binaries to investigate binarity in the WDS catalog; any angular separation that would correspond to $<1\,$AU would be flagged as a binary. However, all of the binaries in our sample that were listed in the WDS catalog were $>10\,$AU.

Of the 180 stars, four were identified as RS CVns: HD 217344, TY Col, AI Lep, and V1198 Ori. An additional 23 stars were identified as having orbital periods $<100$ days, for a total of 27 binaries that might be tidally interacting, reducing our sample size to 153 targets. It is worth emphasizing that although this work ruled out several definitive binary stars, it is notoriously difficult to definitely rule out binarity and there very well could still be binaries contaminating the sample. The final collection of \emph{TESS} data used in this paper can be found in MAST: \dataset[10.17909/8g1x-8945]{http://dx.doi.org/10.17909/8g1x-8945}.

\section{Methods and Results}\label{sec:methods}
\citet{Tu2020} reports the occurrence time and duration of the flares so that we could implement our own energy calculation method for each flare as outlined in Appendix \ref{app:colors}. By contrast, \cite{Doyle2020} does not report information for individual flares. Because of the discrepancies in how the two teams calculated flare energy and reported flare information, and because consistency in flare energy estimation is important for the purposes of our analysis, we produced our own flare-identification and analysis pipeline to run on all 153 assumed single stars that exist in a VLASS field presented by the two catalogs. In this section, we briefly review our flare-identification process (more details in Appendix \ref{app:flare_characterization}) as well as our VLASS source identification method.

\subsection{TESS Flare Identification and Characterization}\label{subsec:Flare_identification}
Our method of identifying flares in \emph{TESS} Science Processing Operations Center (SPOC) light curves closely follows the methodology outlined in Section 3.1 of \citet{Jackman2021}. In this method, stellar light curves are iteratively de-trended by applying a median filter to data and flagging points $>3\sigma$ in order to produce a light curve that has had all flares and eclipses removed. The raw light curve is then divided by this flagged light curve to remove the effects of \textit{e.g.} starspots on the flux. Flare candidates are identified in this de-trended light curve and then checked by eye to confirm they are flares. We provide a more in-depth explanation for this method, the flare-energy calculation, and the flare-rate calculation in Appendix \ref{app:flare_characterization}. The code we wrote and used for this analysis is available for use on GitHub\footnote{\url{https://github.com/iveydavis/flare_id/}}.

\begin{figure}
    \centering
    \includegraphics[width=0.8\linewidth]{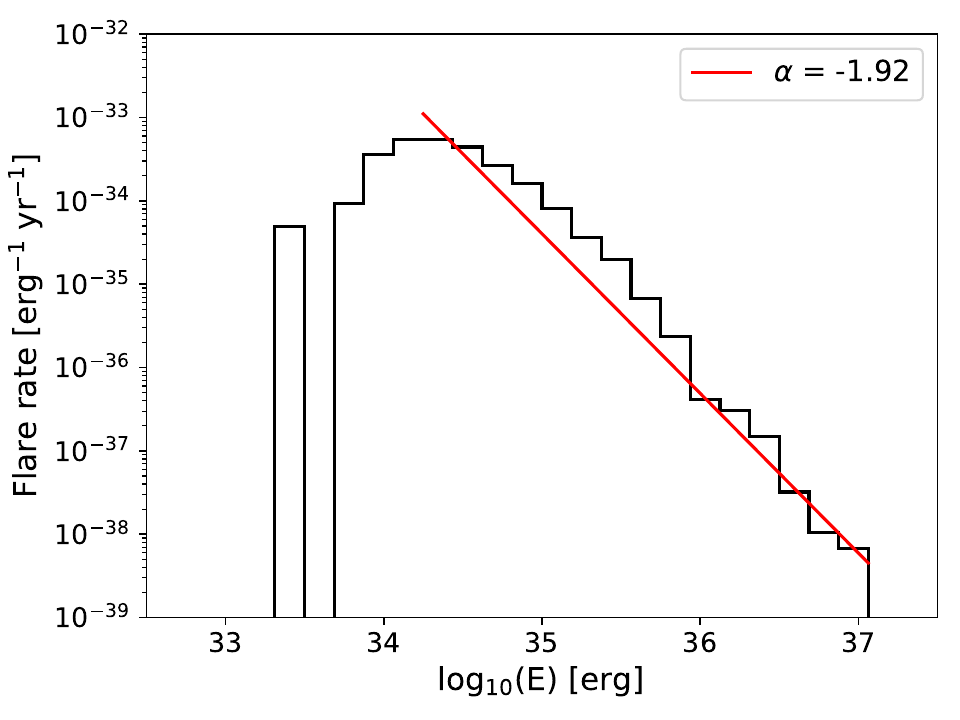}
    \caption{The FFD (black) for the 1,493 flares we identified across 150 solar-type stars. The FFD for flares above $10^{34}$\,erg is best fit by a power law index of -1.92 (red). \label{fig:FFD}}
\end{figure}

Across the 153 stars and all currently completed \emph{TESS} sectors (up to sector 65), the code identified 1,579 flares, 86 of which, we concluded, were not real flares after checking each flare by eye. The total number of flares is then 1,493. Although we did not deliberately select for flares $>10^{33}$ \,erg as \citet{Tu2020} did, our flare-selection criteria resulted in a minimum energy of $\approx10^{33}\,$erg to be detected on stars $\gtrsim5000\,$K and therefore all of our flares qualify as superflares. As a rudimentary validation of our method, we look at the flare frequency distribution (FFD) for the flare energies we calculated (see Figure \ref{fig:FFD}). The slope of the FFD for flares with energies $>10^{34}$\,erg follows a power law index of $\approx-1.9$, which is consistent with other studies of solar-type stars by \textit{e.g.} \citet{Maehara2012},\citet{Shibata2013}, and \citet{Maehara2015}.

Of the 153 stars, there were three for which our flare-finding code found no flares: TIC 328349131 (TYC 58-209-1), TIC 1258935 (TYC 6027-806-1), and TIC 57719552 (TYC 7216-55-1). For each of these stars, we inspected the entire \emph{TESS} light curve by eye. There is a possible flare for both TIC 328349131 and TIC 1258935, but these flares occur at the very end of a light curve segment we had flagged. For TIC 57719552, there is a flare candidate peaking at 1589.347655 [BTJD], but it had an insufficient number of 2.5$\sigma$ points to be identified as a flare. Although the flare candidates for these stars may very well be actual flares, we do not find it appropriate to find all flares that may have been missed across the other 150 stars. Therefore, we exclude these three stars from our sample. Our final sample to look for bursts in VLASS is then 150 stars.

\subsection{VLASS Burst Identification}\label{subsec:vlass_method}
For each of the 150 stars that lie in a VLASS field, we downloaded $2'\times2'$ cutouts of VLASS frames for both E1 and E2 from the CIRADA Cutout Image Web Service. Sources were identified using the \texttt{findsource} function in \texttt{CASA} \citep{CASA2022}. Sources less than 4$\sigma$ were excluded, where $\sigma$ was determined by the root mean square (rms) property of the \texttt{statistics} function applied to the CIRADA frame. For each frame, we used the observation date from the CIRADA cutout header and proper motion values from Gaia eDR3 \citep{Gaia2020} to calculate the proper-motion corrected position of the star that should be in the frame.

While VLASS has $2.5"$ spatial resolution, the quick-look images from CIRADA have a pixel size of $1"$, similar to the survey's maximum positional uncertainty $\approx1"$ (\textit{e.g.} VLASS memos 14\footnote{ \url{https://library.nrao.edu/public/memos/vla/vlass/VLASS_014.pdf}} and 17\footnote{ \url{https://library.nrao.edu/public/memos/vla/vlass/VLASS_017.pdf}}). Because of this, VLASS sources reported in \texttt{findsource} that were $\diff\theta>1"$ from the star's Gaia position were considered emission not associated with the star. CIRADA frames where there was a source coincident with the star's position were then inspected to confirm convergence of a 2D-gaussian point-source model at the location of the source using \texttt{CASA}'s \texttt{imfit} function. We expect these sources to have a root mean square (rms) positional error of $\theta_\text{rms} = \theta_\text{FWHM}\text{SNR}^{-1}(2\ln{2})^{-1/2}$ where $\theta_{FWHM}=2.5"$ is the beam size and SNR is the signal-to-noise ratio of the source \citep{Condon1997}. All of our sources have $\diff\theta<1.5\sigma_\text{SNR}$, making us reasonably certain the radio emission is coincident with the associated star's position.

\begin{figure*}
    \centering
    \includegraphics[width = \textwidth]{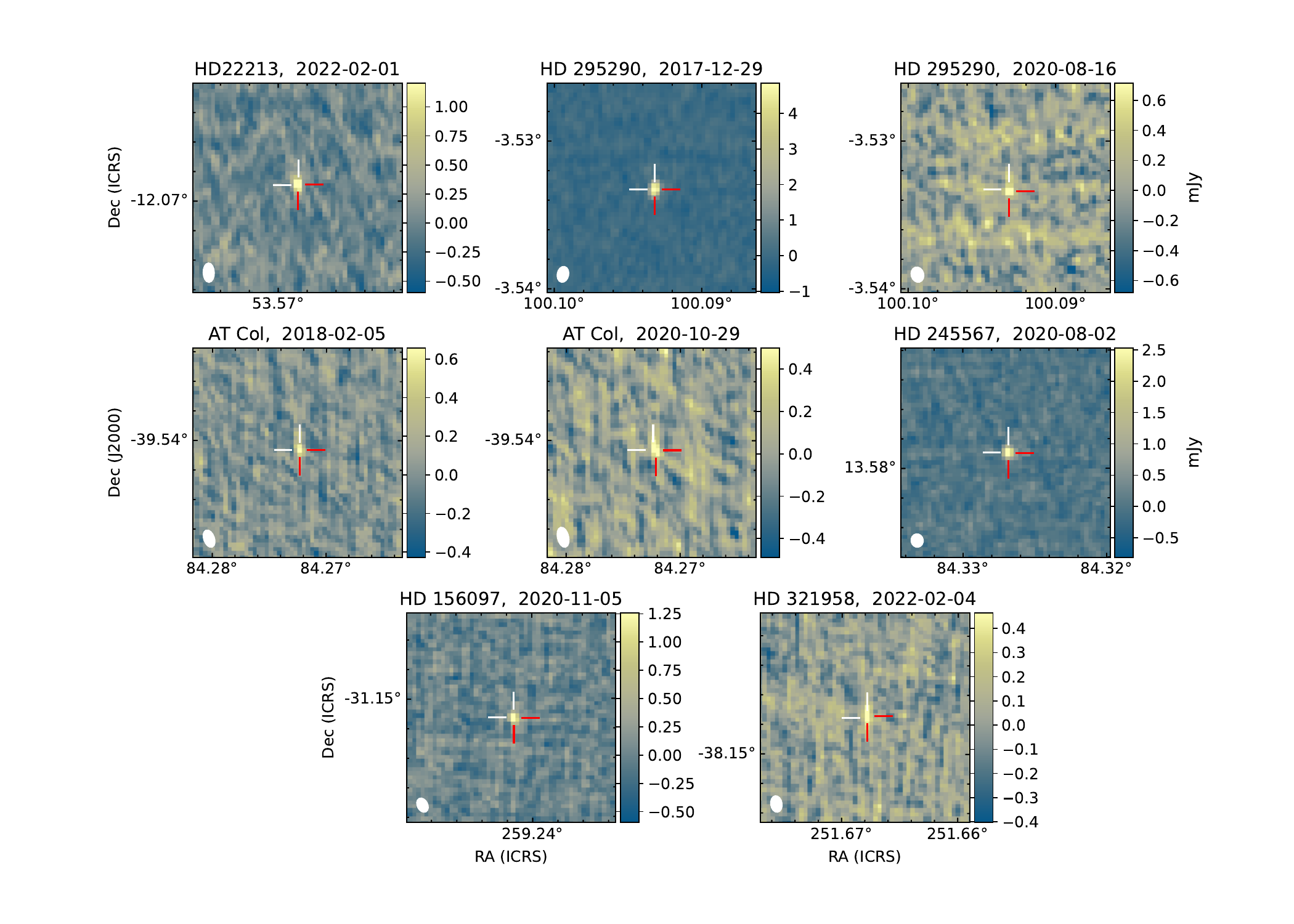}
    \caption{The CIRADA Stokes I cutout for each VLASS source associated with a super-flaring star. The white cross-hairs are centered on the proper-motion corrected position of the star from SIMBAD values \citep{Wenger2000} and the red cross hairs are centered on the position of the source as identified by \texttt{CASA}'s \texttt{imfit} function. The beam shape is shown as a white ellipse in the bottom left corner of each frame.}
    \label{fig:all_detections}
\end{figure*}

This method found eight VLASS sources coincident with the position of six stars presented in \citet{Tu2020}'s and \citet{Doyle2020}'s flare catalogs. Two of these sources (HD 245567 and HD 156097) are the ones previously identified by Ayala et al. (in preparation). The four other stars were excluded from the previous works because they  showed up in E1 but not E2, and/or they didn't meet the $7\sigma$ threshold in E2. Two of these four stars (HD 295290 and AT Col) are detected in VLASS epochs 1 and 2. The VLASS fields for all detected sources are shown in Figure \ref{fig:all_detections}. Details on these detections are provided in Table \ref{tab:radio_info}. 

Of the six stars, two (HD 156097 and HD 321958) were flagged by \citet{Tu2020} for possibly having an M-dwarf within their \emph{TESS} point-spread function (PSF). While the VLASS beam size makes us confident that the  radio emission originated from the solar-type star, there was some risk of the optical flares actually originating from the M-dwarf because of how large the \emph{TESS} PSF is. However, based on the implied flare energies and associated flux contribution that might be expected from an M-dwarf we are able to rule out that an M-dwarf is responsible for the flares. Details on this analysis are provided in Appendix \ref{app:mdwarf_contamination}.

\begin{table*}[t!]

\centering
\begin{tabular}{lccccc|ccccc}
\hline \hline
Star &E1 $S_{\nu, \text{I}}$ & $\log_{10}L_\text{R}$ & $\diff\theta$ & $S_{\nu, \text{V}}$ & $P_\text{V}$ &E2 $S_{\nu, \text{I}}$ & $\log_{10}L_\text{R}$ &$\diff\theta$ & $S_{\nu, \text{V}}$ & $P_\text{V}$\\
& mJy  & erg/s/Hz& $''$ &mJy &  &mJy  & erg/s/Hz& $''$ &mJy & \\\hline
HD 22213 & $<$0.88 & $<$15.44& ---  &--- & --- &1.56 (9.8) &15.69& 0.13 &0.43 (2.8) & 0.28 (2.7)\\
HD 295290 & 6.19 (31.6) &16.43& 0.05 &-0.32 (1.7) &  -0.05 (1.7)&0.85 (4.1) &15.57& 0.35 &-0.44 (2.2) & -0.52 (1.9)\\
AT Col & 0.73 (5.9) &15.70& 0.04 &-0.37 (3.3) &  -0.51 (2.9)&0.612 (4.3) &15.62& 0.70 &-0.21 (1.4) &  -0.34 (1.3)\\
HD 245567 & $<0.48$ &$<$15.81& --- &--- & ---&3.40 (20.4) &16.67& 0.04 &2.07 (12.8) & 0.61 (10.8)\\
HD 156097 & $<0.48$ &$<$15.88& --- & ---& ---&1.57 (10.7) &16.40& 0.23 &-0.15 (1.0) & -0.10 (1.0)\\
HD 321958 & $<0.48$ &$<$16.23& --- & ---& ---&0.54 (4.5) &16.285& 0.51 &0.24 (2.1) & 0.44 (1.9)\\
Median Star & $<0.56$ & $<$16.00& --- & ---& ---& $<$ 0.58 & $<$16.01& --- & ---& ---\\
\hline
\end{tabular}
\caption{Details for the VLASS detections of the six stars in E1 (left) and E2 (right). In addition to the Stokes I flux density ($S_{\nu, \text{I}}$), we also calculate the radio luminosity $\log_{10}L_\text{R}$, the offset between the \texttt{findsource} source position and the Gaia proper motion corrected position ($\diff\theta$), the Stokes V flux density ($S_{\nu, \text{V}}$) for sources detected in total intensity, and the circular polarization fraction $P_\text{V}$. The signal-to-noise ratios for the flux densities and polarization fractions are reported in parentheses and non-detections are reported as $<4\sigma$. The row for Median Star refers to the median values of the \emph{TESS} stars that were not detected in VLASS.}
\label{tab:radio_info}
\end{table*}

\subsubsection{VLASS Polarization}\label{sec:vlass_pol}
For each of the VLASS sources, we re-imaged the field in full polarization. We restored the VLASS observations from the archive using the default VLA Calibration Pipeline for that observation. The sources were imaged in CASA (https://casa.nrao.edu; \citet{CASA2022}) using CASA version 5.6.1-8. We created a wide-band (1965 - 4013 MHz) multi-frequency synthesis image using a single Taylor expansion term representing the average over the band in Stokes I and V. The flux densities in Stokes V polarization was extracted at the location of the peak in the Stokes I image, taking the extremal value (positive or negative) in the V image in a 5 $\times$ 5 pixel box ($2.5"$ $\times$ $2.5"$) around the I peak.

The values reported were those extracted values, with the uncertainties derived using the standard deviation about the mean in the given image calculated in a larger box nearby the source. The Stokes V data have not been corrected for any systematic effects such as the beam squint between the R and L polarized primary beams of the VLA, or any systematic offset between the R and L antenna based calibrations. The VLASS images are made using data from 128 individual on-the-fly (OTF) pointings surrounding the target and we expect some of these effects to average out. Systematic errors in Stokes V are estimated to be $10\%$, therefore circularly polarized emission above the $10\%$ level are likely believable. The flux densities themselves derived from these Quick-Look Images are also subject to significant uncertainty (VLASS Memo 13\footnote{ \url{https://library.nrao.edu/public/memos/vla/vlass/VLASS_013.pdf}}) with peak values showing a systematic offset of {-15\%} and scatter of $\pm 8\%$ in flux density.

\section{Discussion}

Because none of the VLASS detections were contemporaneous with \emph{TESS} observations of the respective stars, we cannot unambiguously say that any radio emission is associated with a superflare. Instead, we can investigate the properties of the VLASS-detected stars and how they compare to the properties of the larger sample of super-flaring stars in \emph{TESS} and to the broader population of radio-detected stars. From this, we can explore what that might mean for the likelihood that these bursts are associated with superflares or their correlation with other stellar properties, essentially independent of spectral type. In particular, we consider what the effects of the observing properties of VLASS, maximum flare energy, and the flare rate might have on detecting radio emission from these stars. Before discussing our results in the context of the larger \emph{TESS} superflare star sample, we first discuss each of the individual VLASS-detected stars.

It is worth mentioning that the Rossby number-- the ratio between the stellar rotation period and the convective turnover time-- is normally included in discussions of activity, but we exclude it here. This is because our stellar sample is so restricted (stars with temperatures between $\sim4900-6000\,$K and rotation periods $\lesssim10\,$d), and more specifically because nearly all of the stars have Rossby numbers that would put them in the saturated regime \citep{Wright2011,landin2022}. Because of this, there should be no correlation between it and other activity indicators. After checking our results, we indeed find there is no correlation between the Rossby number and the flare energies, peak luminosities, or rates of the stars. 

\subsection{The VLASS-detected Stars}
\subsubsection{HD 22213}
HD 22213 is one star in our sample that is actually a known binary. However, the orbital separation is on the order of 100\,AU \citep{Mason2001}. As such, the companion is not expected to be strongly impacting the rotation evolution of the star or driving magnetic activity. Instead, its high activity is likely related to its young age; its inclusion in the Tucana-Horologium association suggests it is $\approx45\,$Myr old \citep{Bell2015}.

HD 22213 was detected in X-ray both by ROSAT in both 1991 \citep{Boller2016} and eROSITA in 2020 \citep{Merloni2024}. The flux in the 0.2-2.3\,keV eROSITA band was $4.76\times10^{-12}$mW/m$^2$, corresponding to an X-ray luminosity of $1.5\times10^{30}\,$erg/s. Using the stellar parameters in table \ref{tab:flare_star_info}, this leads to a ratio of X-ray luminosity to bolometric luminosity of $L_\text{X}/L_\text{bol}= 4.8\times10^{-4}$, within the saturated regime. As mentioned previously, this is to be expected for the stars in our VLASS-detected sample, all of which have Rossby numbers that put them in this regime.

\subsubsection{HD 295290}\label{sec:HD295290}
HD 295290 is a member of the 40\,Myr Columba association \citep{Gagne2018} and is one of two stars in our sample detected in both VLASS epochs that we search for emission. Its estimated brightness temperature in E1 was the highest of any detection from our sample ($\approx10^{11}\,$K) and, when paired with a lack of polarization, makes synchrotron a more likely emission mechanism than gyrosynchrotron for the E1 detection, although other processes cannot be ruled out. When considering the nature of the much weaker E2 emission, it is worth acknowledging that VLASS epoch 3 (E3) data is available for this source. In this epoch, there was an $8\sigma$ detection ($1.273\pm0.161\,$mJy) of the source with no detectable polarized emission (Stokes V flux density $=-0.232\pm0.150\,$mJy). 

HD 295290's E1 radio emission being $\approx10\times$ higher flux density than the emission in the other epochs makes it more likely that the emission in E1 is burst-related in nature and possibly associated with a transient event like a flare. It is less certain whether the emission in E2 and E3 are bursts or if they are quiescent in nature. One way to evaluate the likelihood that the emission is a radio burst associated with a flare is to consider the probability of a flare occurring during a VLASS observation. If a flare were only as long as a VLASS exposure time, then the probability of observing a flare would just be:
\begin{equation}\label{eq:P_burst}
    P = 1 - \exp{(-f_\text{fl}\cdot t_\text{epoch}\cdot N_\text{epoch})}
\end{equation}
where $f_\text{fl}$ is the flare rate of the star, $t_\text{epoch}$ is the exposure time per epoch, and $N_\text{epoch}$ is the number of epochs that we search for a flare during. However, as flares last much longer than 5\,s, this would grossly underestimate the probability of observing a flare during a VLASS epoch. Instead, we must scale the argument in the exponent by $t_\text{fl,avg}/t_\text{epoch}$, where $t_\text{fl,avg}$ is the average duration of a flare, such that:
\begin{equation}\label{eq:P_flare}
    P_\text{flare} = 1 - \exp{(-f_\text{fl}\cdot t_\text{fl,avg}\cdot N_\text{epoch})}.
\end{equation} 

For HD 295290, the average flare duration is 1657\,s and the flare rate is 217$\,\text{yr}^{-1}$, so that the probability of HD 295290 flaring during a single VLASS observation is only $P_\text{flare} = 0.011$ and of flaring in all three epochs $P_\text{flare}^3 = 1.45\times10^{-6}$. This is assuming that the \emph{TESS} and VLASS sensitivities are at the same part of the luminosity function for these flares. Instead, it could be  that only one of these VLASS detections is a burst associated with a superflare while the other two epochs are showing quiescent emission. Considering that the emission from E1 was 5-7 times stronger than the emission observed in E2 and E3, it could be that E1 is emission associated with a burst while E2 and E3 are quiescent emission.

The flux density of even the lower intensity E2 emission would be the equivalent of $\approx10^7$ solar flux unit (SFU) burst at the distance from the Sun. This is an order of magnitude brighter than the brightest burst seen from the Sun at these frequencies \citep{Cliver2011}. Conversely, the radio luminosity of this emission is $\sim10-500\times$ the quiescent emission reported for solar type stars by \citet{GudelSchmittBenz1994}, which is already 1000s times the estimated solar quiescent radio luminosity. Although the emission reported here is brighter, it is unclear if it is to such a degree that it can unambiguously be distinguished from quiescent emission. 

While this burst probability discussion does assume that the radio emission is associated with a flare, it could be that the emission is associated with an unrelated magnetospheric process as is discussed in section \ref{subsec:HD245567}. Such strong and consistently-detected emission may also be explained through binary interactions, which are known to be capable of producing bright, persistent, and variable radio emission \citep{Drake1989}.  Alternatively, accretion (or ejection) processes are also capable of producing long-lasting and variable radio emission \citep{Skinner1994, Cesaroni2024}. Such events may be reasonable to expect given HD 295290's young age. Because the VLASS observing cadence is so sparse, and especially because HD 295290 is detected in all three epochs, it is essentially impossible to derive a specific radio emission rate that we could compare against its flare rate. Instead we will have to consider the burst rate of the entire sample as is addressed in section \ref{sec:flare_rate}. Because of this, the relationship between HD 295290's flare activity and its radio emission remains unclear.

\subsubsection{AT Col}
AT Col is the other star detected in both E1 and E2 of VLASS. It was most significantly detected in E1 when it also exhibited detectable ($>3\sigma$) Stokes V emission. The flux density in Stokes I and V suggest a polarization fraction of $0.52\pm0.18$. Such a high circular polarization fraction cannot be explained by synchrotron emission, which is generally much less than $10\%$ (\textit{e.g.}, \citet{Melrose1971, Sazonov1972}). This would then suggest that either gyrosynchrotron or a coherent emission process is responsible for the emission in E1.

Without constraints on the size of the emitting region, it is difficult to use the brightness temperature to constrain the emission process. The brightness temperature of the Stokes I emission, assuming the emitting region is the size of the star ($\approx2\times10^{10}$\,K), may be sufficient to rule out gyrosynchrotron, which is limited to $\lesssim10^9\,$K \citep{DulkMarsh1982}. The plausible coherent emission mechanisms-- plasma emission and ECME-- both can reach brightness temperatures well in excess of $10^{12}$K and circular polarization fractions in excess of $70\%$, but may have lower polarization fractions due to propagation effects in the stellar atmosphere. Especially because the emission is not strong enough for us to extract a spectrum, we can not conclude whether the emission is gyrosynchrotron or a coherent emission process.

Unlike HD 295290, AT Col is not also detected in E3. This might make it more likely that the emission from AT Col is indeed transient in nature despite the low probability of detecting a flare-associated burst in two epochs ($P_\text{flare}^2\approx10^{-5}$). In a similar argument to the case of HD 295290 in section \ref{sec:HD295290}, the flux density compared to solar radio bursts at these frequencies may be sufficient on its own to rule out quiescent emission, even for the weaker detection which would have a corresponding flux density of $\approx10^7$\,SFU at 1\,AU. While considering the intensity in the context of the Sun might imply that this emission is transient, it is only 1-2 orders of magnitude brighter than quiescent emission from solar-type stars \citep{GudelSchmittBenz1994}; when considering AT Col in the stellar context, it is unclear whether the intensity of its emission can unambiguously rule out quiescent emission.

It is worth noting that AT Col is consistently detected in X-ray missions, including ROSAT in 1990 \citep{ROSAT1995, Mason1995, Boller2016} and 2020 \citep{Merloni2024}, Swift between 2009 and 2016 \citep{Evans2020}, and XMM-Newton in 2014 \citep{Freund2018}. Its X-ray flux from ROSAT implies it is in the X-ray saturation regime $L_\text{X}/L_\text{bol}\approx10^{-3}$ \citep{Vilhu1984, Wright2011}, which again is indicative of it being a highly active star. It may also have an activity cycle, as suggested by its long-term rotational modulation and brightness variability \citep{Distefano2017}. The persistent X-ray emission paired with the activity cycle might explain the detection in E2 and non-detection in E3; AT Col may have been at a cycle minimum during E3, reducing its possible quiescent radio emission to an undetectable level for VLASS. 

\subsubsection{HD 245567}\label{subsec:HD245567}
\begin{figure}
    \centering
    \includegraphics[width=0.8\linewidth]{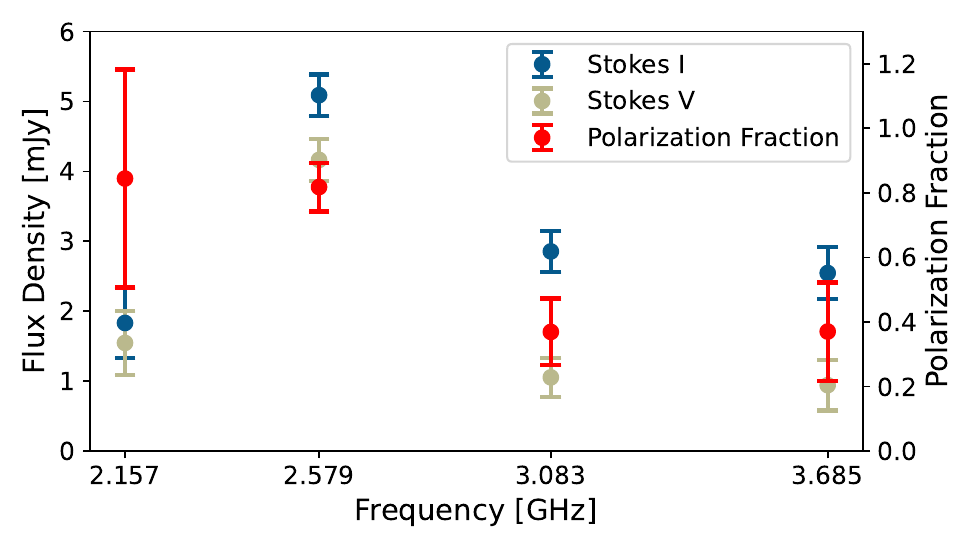}
    \caption{The spectra for Stokes I and V (left axis) and the polarization fraction (right axis) for HD 245567 in VLASS E2. At the peak flux density, the polarization fraction is $0.82\pm0.076$.}
    \label{fig:HD245567_spectra}
\end{figure}

HD 245567, a weak-line T Tauri star \citep{Li1998}, is one of two stars presented here that was identified in the work of Ayala et al. (in preparation). Because it is a weak-lined T Tauri, it is likely that its inner disk has mostly dissipated and therefore does not play a significant role in HD 245567's activity. HD 245567 is also one of the two stars in this work that has significantly circular polarized radio emission. The high flux density-- the second highest of our sample-- made it possible to extract spectra for both Stokes I and V, shown in Figure \ref{fig:HD245567_spectra}. The exceptionally high polarization fraction-- more than $80\%$ in some bands-- requires this emission be coherent. The significance of this in the context of the larger superflare star sample is discussed briefly in sections \ref{sec:gudel_benz} and \ref{sec:flare_rate}.

The two main coherent emission mechanisms for stellar radio emission are plasma emission and ECME, both of which can reach polarization fractions up to 100$\%$ (\textit{e.g.} \citet{Dulk1985}).  Plasma emission from the Sun is never seen to have such a high polarization fraction above a few hundred MHz. Instead, this emission is more likely to be due to ECME. This is observed from the Sun when electrons are trapped in post-flare loops \citep{Gergely1986, Morosan2019} and is also observed from the magnetized planets. In particular, the cadence and geometry of Jupiter's ECME depends on both the orbital phase of its moons \citep{Goldreich1969} and the rotational phase of Jupiter \citep{Zarka2021}.  Because of this, it is unclear whether this emission is associated with flaring behavior or if it is rotation-modulated emission.

The frequency that ECME is observed at depends explicitly on the magnetic field strength of the source with frequency $\nu=s\,2.8\,B$\,[MHz] where $B$ is the magnetic field strength in Gauss and $s$ is the harmonic of the emission we are observing. Assuming this emission is ECME at the fundamental, the detection at 3.685\,GHz would imply that HD 245567 has a magnetic field $\gtrsim1,300\,$G. This field strength is consistent with Zeeman-broadening measurements of T-Tauri magnetic fields reported by \textit{e.g.} \citet{Yang2005,Lavail2017}.

\subsubsection{HD 156097}
HD 156097 was the VLASS-detected star with the highest flare rate and is one of four stars detected only in VLASS E2. It was one of two stars flagged by \citet{Tu2020} as having an M-dwarf contaminating its \emph{TESS} PSF that may be responsible for the flares in the \emph{TESS} light curves. We address this possible contamination in appendix \ref{app:mdwarf_contamination} and find that the nearest star with a temperature reported by Gaia DR3 that is within $42"$ (two pixels on the \emph{TESS} detector) could not reasonably produce flares large enough to account for the flare amplitudes observed from this star. 

HD 156097's kinematics lead the Banyan moving group association program to classify it as a field star with a probability $\approx74\%$\citep{Gagne2018}. If this star is indeed a field star, it may be more likely that a companion-- rather than a young age-- is responsible for its $\approx2\,$day rotation period. The other possible association is the  Upper Centaurus Lupus (UCL) association at 23.5\% probability. This association, estimated to have an age between 10-25\,Myr \citep{Roser2018}, would be consistent with the $\approx30$\,Myr age assumed from its Lithium equivalent width \citep{Desidera2015}.

This star has additionally been detected in X-ray in 1990 \citep{Boller2016}, 2007 \citep{Freund2018}, and 2020 \citep{Merloni2024}. Its flux in the eROSITA 0.2-2.3\,keV band in 2020 was $4.35\times10^{-12}\,$ mW/m$^2$, or an equivalent luminosity of $6.9\times10^{30}\,$erg/s \citep{Merloni2024}. Using the effective temperature and radius for the star reported in table \ref{tab:flare_star_info} to calculate the bolometric luminosity, this would put it at $L_\text{X}/L_\text{bol}= 8.5\times10^{-4}\approx 10^{-3}$, which implies it is in or near the saturated regime. This is consistent with the star being highly active.

\subsubsection{HD 321958}
Like HD 156097, HD 321958 was flagged by \citet{Tu2020} to have an M-dwarf in its \emph{TESS} PSF. However, the flare luminosity that would be required from the M-dwarf to reproduce the flares observed in the \emph{TESS} light curve would be orders of magnitude higher than the quiescent luminosity of the M-dwarf, making it a highly unlikely contaminant in our analysis. This star was marginally detected only in VLASS E2, but has the third highest radio luminosity and the largest flare energy of the VLASS-detected stars. These details are likely observationally driven as it is the furthest star of the VLASS-detected stars.

Banyan associates HD 321958 with the UCL moving group with a high probability ($85\%$), with a small probability ($15\%$) that it is a field star. It is in fact apart of the defining sample for the UCL moving group \citep{Roser2018}. It was also observed in the X-ray in 1990 by ROSAT \citep{Boller2016} and in 2020 by eROSITA \citep{Merloni2024}. Its flux in the 0.2-2.3\,keV band measured by eROSITA is $9.18\times10^{-13}$mW/m$^2$. When paired with the bolometric luminosity estimated from the stellar parameters provided in table \ref{tab:flare_star_info}, this leads to $L_\text{X}/L_\text{bol}= 6.0\times10^{-4}$, within the saturated regime as would be expected for a star this young.

\subsection{VLASS detection limits}\label{subsec:detect_lims}
If the stars VLASS detected had systematically low radio luminosities due to being especially nearby or in low-noise VLASS fields, it would make sense that we were able to detect them over the other 144 stars in the \emph{TESS} superflare sample. To investigate this possible bias, we calculate how large the radio luminosity from each of the 144 stars would have to be in order to be a $4\sigma$ detection in their respective CIRADA frame.  This luminosity is taken to be  $4\sigma\times(4\pi\,d^2)$, where $d$ is the distance to the star and $\sigma$ is the rms value for the CIRADA frame as given by \texttt{CASA}'s \texttt{statistics} tool. The median value for the minimum required radio luminosity for the \emph{TESS} superflare stars is reported in Table \ref{tab:radio_info} along with the actual radio luminosities of the six VLASS-detected stars. 

What is immediately obvious is that the radio luminosity of the VLASS-detected stars lie on either side of the median value for the required luminosity for a star to be detected in VLASS. This implies that the majority of the required radio luminosities should not be considered unreasonably large for fast-rotating, solar-type stars and, therefore, the sensitivity of VLASS likely is not an inhibiting factor in this science. That said, it may be that VLASS is only sensitive to bursts associated with the most energetic flares, as is discussed in more detail in section \ref{sec:flare_energy}.

We can also address this from the perspective of the solar paradigm if we  explicitly consider the radio luminosities of the stars detected in VLASS. Five out of six of these stars would have corresponding flux densities of $\approx10^7$\, solar flux units (SFU $=10^4\,$Jy) at 1\,AU. This is only about a factor of 10 brighter than the brightest solar burst observed at similar frequencies \citep{Gary2019, Cliver2011}. When considering that young stars tend to have both denser plasmas and stronger magnetic fields-- thus providing a larger high-energy electron population for emission-- these luminosities seem reasonable in the solar paradigm, granted these exceptionally bright solar radio bursts at 1.4\,GHz do not behave like typical solar radio bursts. The question then is whether there are features of the stars besides their location that might make them more or less likely to be detected in VLASS. This is addressed in the next three subsections.

\subsection{Flare Luminosity and the G\"udel-Benz Relation}\label{sec:gudel_benz}
Although the G\"udel-Benz relation had been derived for the steady, quiescent X-ray and radio emission for binaries, M-dwarfs, and solar-type stars, it has been shown to also apply to the X-ray and microwave emission of both solar and stellar flares \citep{Gudel1993, BenzGudel1994, GudelSchmittBenz1994}. Because of this relation, we might be able to better constrain whether these bursts are associated with superflares by comparing the radio luminosity of the VLASS bursts to the X-ray luminosities of flares from these stars.

To do this, we take $L_x$ to be $10\%$ of the average peak bolometric luminosity of the superflares of a given star (our bolometric luminosity estimate is described in part in appendices \ref{app:colors} and \ref{app:mdwarf_contamination}). We believe this is a reasonable order-of-magnitude estimate given that $\sim70\%$ of the bolometric energy of solar flares is expected to be contained in the white light component of the flare, the flare component that is measured by \emph{TESS} \citep{Kretzschmar2011}. This then leaves on order $10\%$ of the energy to be radiated in other parts of the EM spectrum, primarily the X-ray. We calculate $L_\text{R} = S_\nu\cdot4\pi d^2$ from the flux density of the VLASS sources $S_\nu$ and distance of the star $d$.

\begin{figure}
    \centering
    \includegraphics[width =0.8\linewidth]{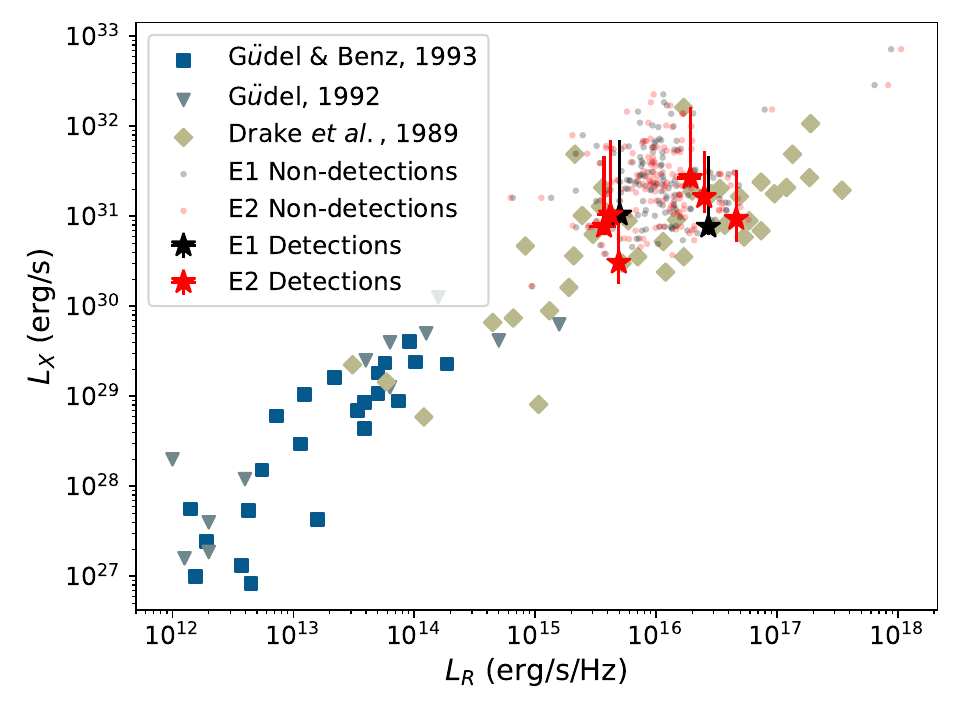}
    \caption{Estimated $L_\text{X}$ plotted against $L_\text{R}$ for the six VLASS-detected superflaring stars as well as three other populations of stars. The extent of the vertical error bars represent the peak luminosity for the brightest and dimmest flares of the star and the horizontal error bars are calculated from the noise in the CIRADA frame. The minimum required radio luminosity for a $4\sigma$ detection vs. the estimated X-ray luminosity for each of the 144 other \emph{TESS} superflaring stars are also included as points.}
    \label{fig:gudel_benz}
\end{figure}

We plot $L_x$  against $L_\text{R}$ for each of the six VLASS-detected stars along with three groups of stars that informed the G\"udel-Benz relationship \citep{Gudel1992,Gudel1993,Drake1989} in Figure \ref{fig:gudel_benz}.  The six VLASS-detected stars follow the $L_x \approx 10^{15.5}\times L_\text{R}$ well within error bars, suggesting the radio emission may indeed have a coronal origin associated with a superflare. We also include the population of \emph{TESS} superflaring stars that were not detected in VLASS in this plot, using $10\%$ of the peak luminosity of the brightest flare from each star as $L_x$ and using the required detection luminosity described in section \ref{subsec:detect_lims} as $L_\text{R}$. Interestingly, these values are also consistent with the G\"udel-Benz relation. This might suggest that should a radio burst associated with a superflare occur on any of the stars in our sample, it would have been detectable in VLASS.

In this context, it is worthwhile again to single out the coherently-emitting HD 245567. Because its emission is unambiguously coherent, we should not expect its radio burst luminosity to be correlated with flare luminosity in the G\"udel-Benz framework. It could be entirely serendipitous that its estimated X-ray luminosity aligns it so well with the G\"udel-Benz relation. Alternatively, there have been reports of coherent radio emission-- albeit at much lower frequencies-- following the G\"udel-Benz relation \citep{Vedantham2022}. It is also worth emphasizing again the likely high errors associated with our flare luminosity calculation given that we do not actually know the temperature of the flares. Without better constraints on the flare properties or contemporaneous observations of this emission in the radio and X-ray, discussion of what this means for the underlying physics responsible for the empirically-derived G\"udel-Benz relationship is outside of the scope of this paper.

\subsection{Flare energy}\label{sec:flare_energy}
\begin{figure*}
    \centering
    \includegraphics[width=0.8\linewidth]{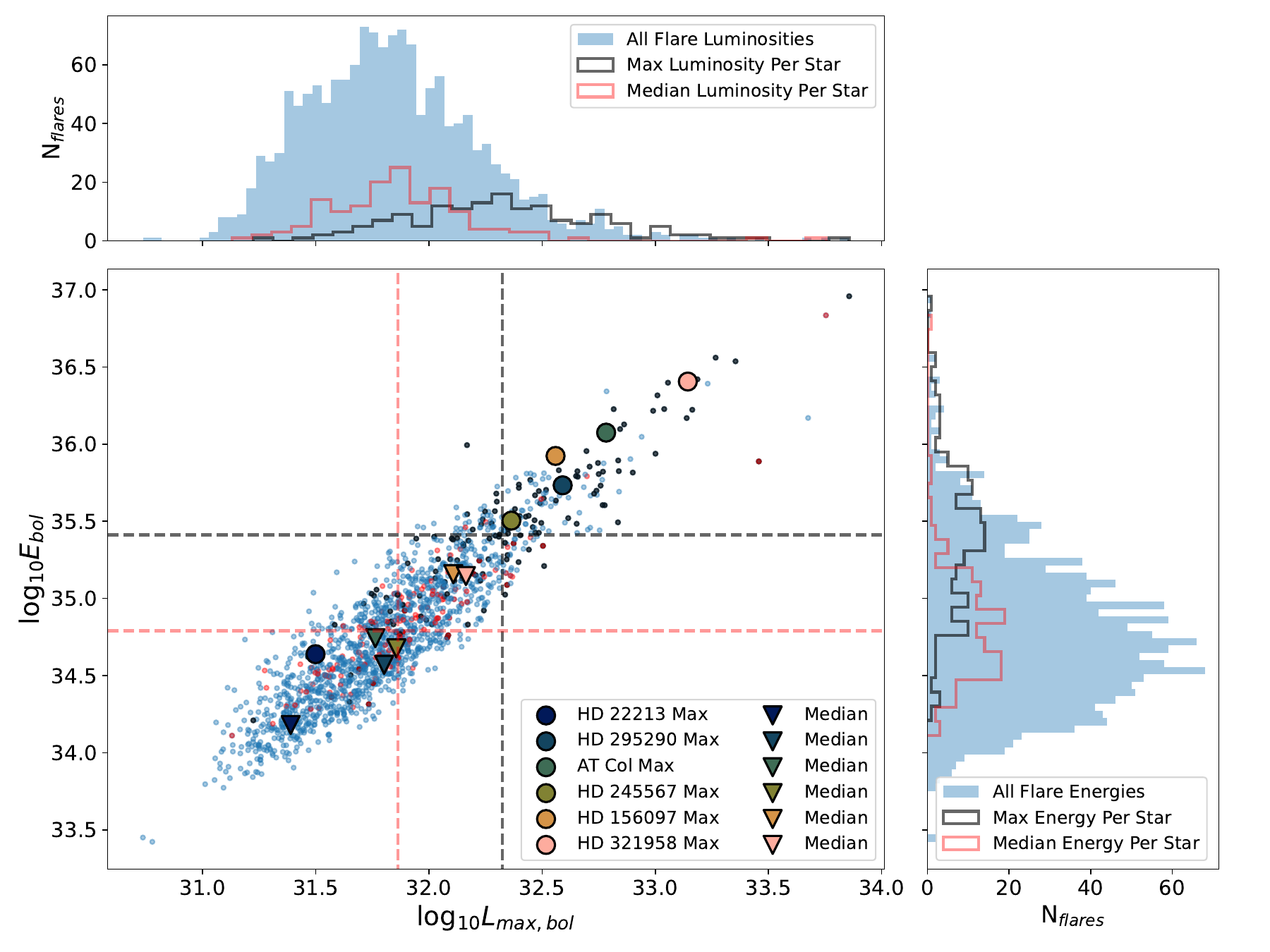}
    \caption{The bolometric energy plotted against the peak luminosity for each flare we identified (blue circles). We also include the median flare (red) value and maximum flare values (black) for each star. Histograms for all flares, median flare per star, and peak flare per star are provided in the top and right panels of the figure in the same colors, and the median values are indicated by dashed lines in their respective colors. The max (circles) and median (triangles) flare values for each of the six VLASS-detected stars are also plotted.}
    \label{fig:lum_v_energy}
\end{figure*}

We see that some flare-related phenomena are more likely to occur on the Sun if certain thresholds are exceeded; for instance, \citet{Yashiro2006FlareCMErelation} report essentially a one-to-one correlation between solar flare energies higher than $10^{30}\,$erg and the occurrence of a CME. This prompts us to investigate whether the the VLASS-detected stars are exceptional in this respect. The flare energy is plotted against the peak luminosity of the flare in Figure \ref{fig:lum_v_energy} for all flares we identified in \emph{TESS}.

If we consider the median flare energy for each star, then the VLASS-identified stars do not seem extraordinary; they lie on either side of the median value of the median flare energy for all \emph{TESS} superflaring stars. However, all of the stars detected in VLASS, except for HD 22213, had maximum flare energies higher than the median value for maximum flare energy for the \emph{TESS} superflare star sample. This might suggest then that radio bursts detectable by VLASS occur preferentially on stars that can produce more energetic flares. By extension, it may be that such bursts are only detectable in VLASS when they are associated with the most energetic flares. This is consistent with the idea that the radio emission is associated with superflares, but the part of the flare luminosity function sampled by VLASS is different from what is sampled by TESS. 

\subsection{Flare rate}\label{sec:flare_rate}
If the radio emission from these stars is associated with a superflare then it would make sense that the more time the star spends flaring, the more likely we would be to detect it in VLASS (assuming VLASS and \emph{TESS} probe the same part of the flare luminosity function). The flare rates for the 150 stars with superflares in \emph{TESS} are calculated using the method described in Appendix \ref{app:flare_characterization} and the median flare rate, as well as the flare rates for the six VLASS-detected stars, are reported in table \ref{tab:flare_star_info}. All of the VLASS-detected stars have flare rates above the median value of the 150-star sample, consistent with the idea that VLASS may have been able to detect radio emission from them simply because they were flaring more often. It is interesting to note that the VLASS-detected star with the lowest flare rate is HD 245567 which, as described in section \ref{subsec:HD245567}, could be producing radio emission through a mechanism other than flares.

Although the VLASS-detected stars flare more than the typical star in the sample, we have already briefly discussed in section \ref{sec:HD295290} how unlikely it is for a flare from one of the VLASS-detected stars to occur during a VLASS observation. The VLASS-detected star with the highest flare rate is HD 156097 at 246 flares per year and an average flare duration of 2,492\,s. Using these values in equation \ref{eq:P_flare}, this makes the probability of seeing it in $N_\text{epoch}=1$\,epoch 0.019, and the probability of seeing it in at least one of $N_\text{epoch}=2$ epochs is 0.038. Conversely, the probability of seeing the median star flare during at least one of two VLASS epochs is only 0.006. Because of this, the exceptionally low probability of observing a flare during any single VLASS epoch may be sufficient to explain why only six superflaring stars were observed in VLASS.

Although the VLASS-detected stars were more likely to be flaring during a VLASS observation, the probability of making a radio detection of even our most frequently-flaring star is low. The question then is how this probability compares to the probability of making a radio detection of a star independent of flare rates. Because the VLASS exposures are so brief and infrequent for any individual source, it is difficult to derive a meaningful radio burst rate for a specific star. Instead, we consider the ensemble radio burst rate of super-flaring stars observed in \emph{TESS}, $f_\text{burst}$. This is given as:
\begin{equation}
    f_\text{burst} = \frac{N_\text{burst}}{N_{\star}\cdot t_\text{epoch}\cdot N_\text{epoch}} \:[\text{burst}\cdot \text{star}^{-1}\cdot\text{s}^{-1}]
\end{equation}
where $N_\text{burst} = 8$ is the total number of radio bursts found (assuming all radio detections are a burst rather than quiescent emission), $N_{\star} = 150$ is the number of \emph{TESS} super-flaring stars we looked for radio emission from, $t_\text{epoch} =5\,$s is the exposure time for each star per VLASS epoch, and $N_\text{epoch} = 2$ is the number of VLASS epochs we searched for emission. 

Because we do not know the duration of the bursts, we assume that the duration is the same as $t_\text{epoch}$ so that we can use $f_\text{burst}$ in place of $f_\text{fl}$ in equation \ref{eq:P_burst}. The probability of any super-flaring star being detected in a single  epoch is then 0.026, and of being detected in at least one of $N_\text{epoch} = 2$ epochs is 0.052. Alternatively, if we acknowledge that the HD 245567 detection is coherent emission possibly not associated with a flare, then $N_\text{burst} = 7$ and the probability of detecting a burst in a single epoch is $P_\text{burst}=0.023$. This is relatively consistent with the probability of detecting a flare from HD 156097 and may indicate that these bursts are indeed associated with superflares.

\section{Conclusion}
In this paper, we used catalogs of flaring, solar-type stars from the first year of \emph{TESS} data  by \citet{Tu2020} and \citet{Doyle2020} to act as a sample to look for associated radio emission from VLASS epochs 1 and 2. Of the 150 single stars from these catalogs that were covered in a VLASS field, six were found to have radio emission, two of which had radio emission present in multiple epochs. Given the luminosities of the flares from each star and the sensitivity of VLASS, we find that the detectability of radio emission from the rest of the superflare sample likely is not dependent on the required radio luminosity of a burst from a given star. Instead, likely what limited the number of radio detections was the low observing cadence of VLASS, with stars that flare more frequently and to a higher energy being preferentially detected in the VLASS. This dependence on flare rate and energy, as well as an adherence to the G\"udel-Benz relation by the VLASS-detected stars, suggests that these radio bursts are likely associated with superflares. 

There were three VLASS-detected stars that stood out: HD 245567, which had a circular polarization fraction indicative of a coherent emission mechanism; AT Col, which was detected in both epochs and exhibited significant polarization in one of the epochs; and HD 295290, which was detected in both VLASS epochs systematically searched here in addition to the currently incomplete epoch 3. The coherent emission from HD 245567 is likely ECME, which would imply a magnetic field strength of $\gtrsim1300\,$G. The fact that this emission places it along the G\"udel-Benz relation highlights the necessity for simultaneous radio, optical, and X-ray observations of the star in order to understand both the nature of the emission and the G\"udel-Benz relation. 

Although the highly circularly polarized emission from AT Col is almost certainly associated with a burst, it is unclear whether all VLASS radio emission from it is burst related. For AT Col and HD 295290, the high flux density (associated with radio luminosities $>10\times$ the brightest solar radio burst) of the sources may suggest that each detection is associated with impulsive magnetic activity such as a flare in spite of the low probability of a flare occurring during any single VLASS epoch. Although this emission is a factor of ten brighter than the brightest solar \textit{bursts}, it is also only 10s-100s times brighter than the \textit{quiescent} emission from active solar-type stars \citep{GudelSchmittBenz1994}. And although the radio luminosities associated with their weakest detections (both occurring in E2) are the same order of magnitude as those of the stars detected in only a single VLASS epoch which might then imply that all of the emission observed in VLASS are transient in nature, they are also the weakest of the E2 VLASS detections. Follow-up observations of HD 295290 and AT Col are needed to conclude whether all of their radio emissions are indeed associated with  bursts or if they represent some of the brightest quiescent emission observed from solar-type stars.

This work demonstrates the power of VLASS to identify active solar-type stars. It also emphasizes the necessity of both dedicated and multi-wavelength campaigns of such stars. Having both time series and higher spectral resolution of bursts from the stars presented here would provide the opportunity to distinguish the emission mechanism as well as possibly derive information on the electron population responsible for the emission. While priority for dedicated follow-up should be given to the six stars we identified to have transient radio emission, we predict that any of the super-flaring stars investigated in this work would produce a detectable radio burst. Studying the larger superflare star sample more in-depth at frequencies $\lesssim1\,$GHz may prove instrumental to understanding the particle acceleration process and particle energy distribution for a population meant to represent the young solar environment.

\section{Acknowledgements}
This work was supported by a grant from the Simons Foundation (668346, JPG). This work is based on data taken with the Very Large Array, operated by the National Radio Astronomy Observatory (NRAO). The NRAO is a facility of the National Science Foundation operated under cooperative agreement by Associated Universities, Inc. This research has made use of the CIRADA cutout service at cutouts.cirada.ca, operated by the Canadian Initiative for Radio Astronomy Data Analysis (CIRADA). CIRADA is funded by a grant from the Canada Foundation for Innovation 2017 Innovation Fund (Project 35999), as well as by the Provinces of Ontario, British Columbia, Alberta, Manitoba and Quebec, in collaboration with the National Research Council of Canada, the US National Radio Astronomy Observatory and Australia’s Commonwealth Scientific and Industrial Research Organisation. This research made use of APLpy, an open-source plotting package for Python \citep{Aplpy} as well as Lightkurve, a Python package for \emph{Kepler} and \emph{TESS} data analysis \citep{Lightkurve}.

\appendix

\section{Effects of Color Correction On Flare Energy}\label{app:colors}
In equations 2-4 of \citet{Tu2020}, they report the bolometric flare flux as being:
\begin{equation}
    F_\text{flare} = (F - F_\text{q})4\pi R_\star^2\sigma_\text{SB}T_\star^4
\end{equation}
where $F$ is the normalized light curve flux, $F_\text{q}$ is the normalized quiescent flux of the light curve, $\sigma_\text{SB}$ is the Stefan-Boltzmann constant, and $R_\star$ and $T_\star$ are the effective radius and temperature of the quiescent star respectively. However, this doesn't account for the fact that flares peak much more in the blue than the quiescent stellar surface, which is an important consideration for \emph{TESS}, which has a preferentially red band.

To account for this temperature and color difference, the bolometric flare flux should instead be:
\begin{equation}\label{eq:flare_flux}
    F_\text{flare} = (F - F_\text{q})4\pi R_\star^2\sigma_\text{SB}T_\text{fl}^4\frac{\int_{\text{I}}B(\lambda,T_\text{q})\diff\lambda}{\int_{\text{I}}B(\lambda,T_\text{fl})\diff\lambda}
\end{equation}
where $T_\text{fl}$ is the temperature of the flare and $B(\lambda,T)$ is the Planck function for a blackbody. $B(\lambda,T)$ is integrated over the wavelengths of the \emph{TESS} band (I), between 600\,nm and 1000\,nm.

Using the median quiescent stellar temperature from their sample (5,500\,K) and assuming a flare temperature of 10,000\,K (\textit{e.g.} \citet{Hawley2003,Kowalski2010,Kretzschmar2011}), this then suggests that \citet{Tu2020} might have been underestimating the flare energies by about at least a factor of 2. It has been shown more recently by \citep{Berger2023} that flares may be significantly hotter than the $\sim10,000\,$K that is regularly assumed, and so this underestimate in flare energy may be even more severe.

\citet{Doyle2020} use a different method: they assume the luminosity of the flare is the same luminosity as the quiescent star. The energy would then be calculated by multiplying the luminosity by an effective duration, determined by the area under the light curve of the flare. Even though they do account for the \emph{TESS} bandpass effect on the quiescent flux, because they also do not account for the temperature difference between the quiescent star and the flare this would suggest an underestimate of the flare energy. Indeed, for each star that is common between \citet{Doyle2020} and \citet{Tu2020}'s work, \citet{Doyle2020} reports a lower maximum flare energy than the already underestimated flare energy of \citet{Tu2020}.

\section{Identifying and Characterizing Flares}\label{app:flare_characterization}
As mentioned in section \ref{subsec:Flare_identification}, our method for identifying flares closely follows \citet{Jackman2021}'s method used for \emph{Kepler} data. Our full method is described here.

We used the \texttt{lightkurve} package \citep{Lightkurve} to access light curves from the Science Processing Operations Center (SPOC) pipeline for the 153 stars. We used this package's box-least-squares method to produce a periodogram for each star. The period at maximum power from the periodogram was taken to be the star's rotational period $P$. This period was used to determine the median-filter window size-- the window was taken to be $P/15$. If the window size was less than 100 minutes, then the window size was set to 100 minutes.

Like \citet{Jackman2021}, we split the full light curve into separate continuous segments before smoothing to avoid jumps in measured flux between gaps and entire sectors.  Gaps in data were required to be at least 30 minutes for the two sections to be considered separate segments. Segments that were less than the window size were excluded from the flare search.

After splitting the light curve, each light curve segment was iteratively smoothed. At each iteration, we used \texttt{scipy}'s \citep{2020SciPy-NMeth} \texttt{medfilt} function to produce a median filter of the light curve segment. The segment was then divided by this smoothed curve to produce a normalized curve. Points in the normalized curve that were greater than $3\sigma$ from the median value were masked in the original segment. This was repeated 20 times, or until there were no outlying points. We repeated this process using a window twice the size of the original window to mask points \textit{less} than $3\sigma$ from the median value to remove eclipses.

A final median filter was produced using the final, masked light curve segment. The original, non-masked light curve is then divided by this median light curve. We clipped points within $P/45$ of the edges of the light curve segments. This was done to prevent the code from mistakenly identifying increased flux at the edges (an effect of the median light curve quotient) as a flare. Points greater than $2\sigma$ were flagged as a preliminary flare candidate. We then required that segments: 1.) contain at least three $3\sigma$ points, 2.) contain at least an additional two $2.5\sigma$ points, and 3.) that consecutive points be separated by no more than 8 minutes to allow for some flexibility regarding erroneously masked points. After flares were identified, we consolidated flares where the end of one flare was within 2.4\,hr of the start of the preceding flare so that complex flare events were not treated as multiple individual flares, which would skew the flare rate to be anomalously high.

After the flares were identified and consolidated, we checked all flare candidates to make sure they had the correct morphology (steep rise, exponential decay) to be considered a flare. For exceptional cases-- \textit{e.g.} extremely short rotation periods-- we went back and manually assigned window sizes and re-ran the procedure in order to appropriately detrend the rotational modulation. When this happened, it often meant also identifying when the derived period was not the actual rotation period. These periods were then also manually updated. 

\begin{figure}
    \centering
    \includegraphics[width = 0.9\linewidth]{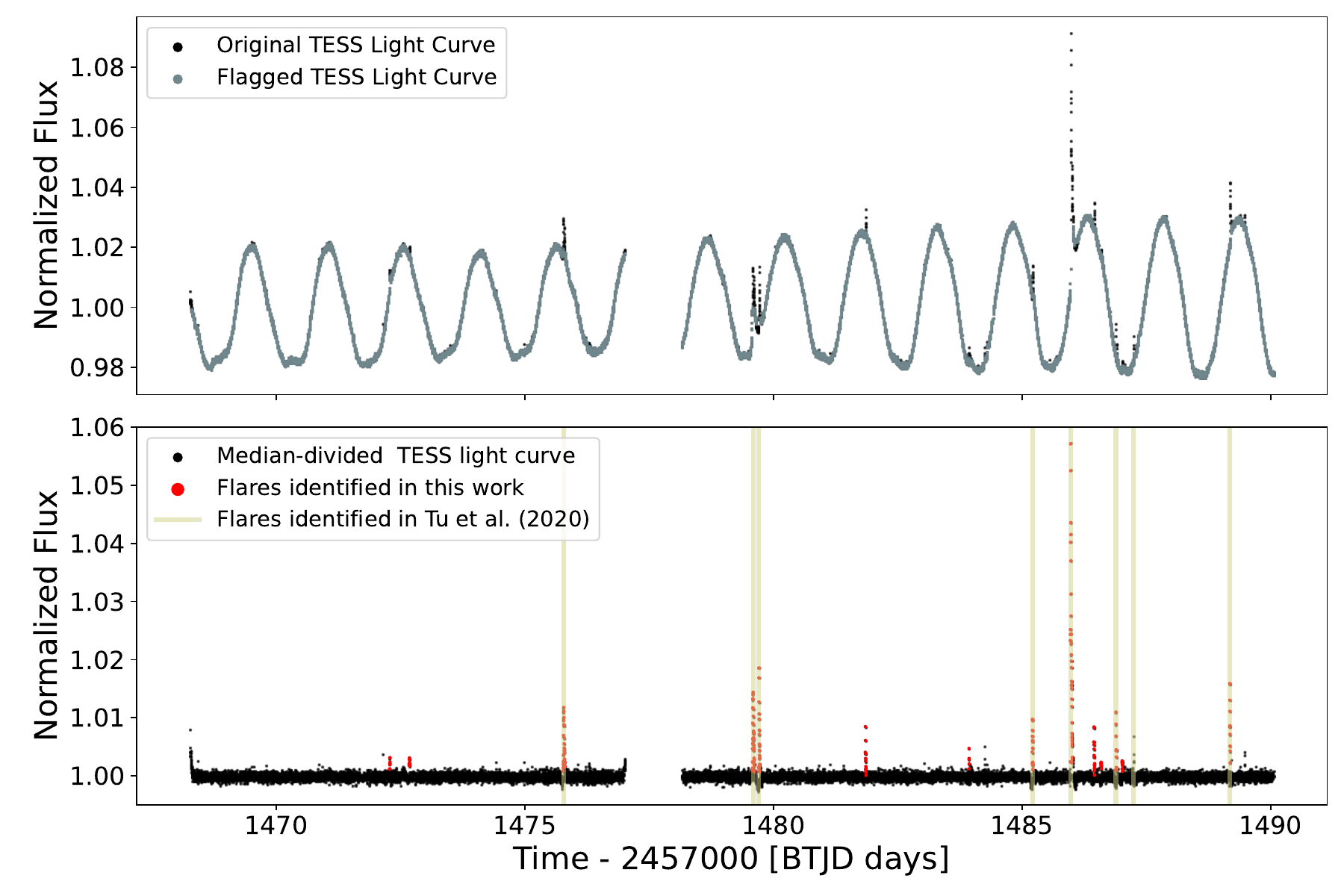}
    \caption{The original SPOC light curve and the light curve with masked outlying points for HD 295290 (top) and the corresponding detrended light curve with the identified flares (bottom). Because \citet{Tu2020} provided the times for the flares they identified, the peak times of their flares are also included in the bottom plot. The flare-identification framework identified 14 flares for HD 295290 compared to \citet{Tu2020}'s 8 flares and \citet{Doyle2020}'s 16 flares. It is possible that many of the flares that we detected that \citet{Tu2020} did not report is because they did not qualify as superflares from their energy calculations and so would be excluded in their analysis.}
    \label{fig:flare_flagging_comparsion}
\end{figure}

The bolometric flux of the flares were calculated using equation \ref{eq:flare_flux} for each of the points in the flare light curve and assuming a flare temperature of 10,000\,K. The bolometric energy is then determined by multiplying each of these fluxes by the surface area of the star and the \emph{TESS} exposure time (120\,s). The flare rate was calculated by dividing the number of flares found by the duration of light curve was searched for flares (\textit{i.e.}, the edges of light curves that were clipped were not included in the flare rate calculation). All data for the 153 stars and their flares are on the github repository along with the code and examples on how to use it. An example of how this flare-finding method performs is shown for HD 295290 in figure \ref{fig:flare_flagging_comparsion}.

\section{Ruling out M-dwarf Contamination}\label{app:mdwarf_contamination}

The concern of an M-dwarf contaminating the \emph{TESS} PSF can be addressed on an energetic basis by assuming the \emph{TESS} light curve for the solar-type star includes the quiescent flux of the star $F_\star$, the quiescent flux of an M-dwarf $F_\text{M}$, and the flux of a flare originating from the M-dwarf $F_\text{fl} = x\,F_\text{M}$. The relationship between the fractional increase in the \emph{TESS}-band flux of the M-dwarf due to a flare and the factor $x$ by which the M-dwarf's flux increases from the flare is then:

\begin{equation}\label{eq:fractional_increase}
    x = \frac{F_\text{N}(F_\star + F_\text{M}) - F_\star}{F_\text{M}} -1 
\end{equation}
where $F_\text{N}$ is the  normalized flux of the light curve (\textit{e.g.} if $F_\text{N} = 1$, the flux is simply from the quiescent star and M-dwarf and $x = 0$). The flux in the \emph{TESS} band from a given star is estimated as:

\begin{equation}
    F = \frac{R^2}{d^2}\int_{600\text{nm}}^{1000\text{nm}}B(\lambda,T)\diff\lambda
\end{equation}
where $R$ is the radius of the star, $d$ is the distance to the star, $T$ is the temperature of the star, and $B(\lambda,T)$ is the Planck function for a blackbody. The bolometric luminosity of the flare on the M-dwarf, $L_\text{fl,bol}$ can then be estimated as:

\begin{equation}\label{eq:flare_luminosity}
    L_\text{flare} = x4\pi R_\star^2\sigma_\text{SB}T_\text{fl}^4\frac{\int_{\text{I}}B(\lambda,T_\text{q})\diff\lambda}{\int_{\text{I}}B(\lambda,T_\text{fl})\diff\lambda}
\end{equation}

\begin{table}
    \centering
    \begin{tabular}{ccccccccccccc}
    \hline \hline 
         &$T_\star$ & $R_\star$  & $d_\star$ &$T_\text{M}$ & $R_\text{M}$ & $d_\text{M}$&$F_\text{N,min}$ & $x_\text{min}$ & $f_\text{min}$ & $F_\text{N,max}$ & $x_\text{max}$ & $f_\text{max}$ \\
        &K & R$_\odot$ & pc & K & R$_\odot$ & pc & {} & {} & {} & {} & {} & {}  \\\hline
        HD 156097 & 5,844 & 1.44 & 116 & 3943 & 1 & 358 &1.003 & 0.29& 0.56& 1.023 & 2.23& 4.27\\
        HD 321958 & 5,626 & 1.26 & 172& 3775 & 1 & 910 & 1.005 & 1.34& 2.47& 1.129 & 34.53& 63.69\\\hline
    \end{tabular}
    \caption{Details on HD 156097, HD 321958, and the M-dwarfs contaminating their \emph{TESS} PSF. $x_\text{min}$ and $x_\text{max}$ are the fractional change in flux in the \emph{TESS} light curve for the dimmest and brightest flare respectively. $f_\text{min}$ and $f_\text{max}$ are the required change in the M-dwarf's bolometric luminosity for $x_\text{min}$ and $x_\text{max}$ respectively. \label{tab:m_dwarf_contamination}}
\end{table}

We use the Gaia DR3 catalog to look for stars within $42"$ (two pixels on the \emph{TESS} detector) of HD 156097 and HD 321958. The most nearby star within HD 156097's \emph{TESS} PSF with a reported temperature is Gaia DR3 5980482341853888384 at a distance of 358.1\,pc and a temperature of 3943\,K. For HD 321958, the most nearby star within its PSF with a reported temperature is the 3775\,K star Gaia DR3 5971100445300717440 at 910\,pc. Both of these contaminating stars have $\log(g) >4$, suggesting they are dwarf stars. For this argument, we favor the contaminants by assuming they have radii $R = 1\text{R}_\odot$ despite their low temperatures and $\log(g)$ values that imply smaller radii. As in Appendix \ref{app:colors}, we take the flare temperature to be 10,000\,K. Using these values in equations \ref{eq:fractional_increase} and \ref{eq:flare_luminosity}, we find that the M-dwarfs contaminating  HD 156097 and HD 321958 would need to reach bolometric luminosities $\sim4\times$ and $\sim64\times$ the quiescent bolometric luminosities of their respective stars in order to replicate the brightest flares observed for these stars. M-dwarf bolometric flare luminosities only exceed the quiescent bolometric luminosity in extreme circumstances \citep{Chang2018}, and have never been reported to reach as high as $17\times$ their quiescent luminosity. Because of this, we find it unlikely that M-dwarf contaminators are responsible for the flares in these light curves. A summary of the stellar values that were used in equations \ref{eq:fractional_increase} and \ref{eq:flare_luminosity} as well as the results of the calculations are given in table \ref{tab:m_dwarf_contamination}.

\bibliography{bibliography}{}
\bibliographystyle{aasjournal}

\end{document}